\documentclass[aps,prb,twocolumn,showpacs,preprintnumbers,amsmath,amssymb,superscriptaddress]{revtex4}%

\usepackage{graphicx}
\usepackage{dcolumn}
\usepackage{bm}
\usepackage{color}
\begin{document}

\preprint{PREPRINT (\today)}

\title{Evidence for 3D-xy critical properties in underdoped YBa$_{2}$Cu$_{3}$O$_{7-\delta }$}

\author{T. Schneider}
\email{toni.schneider@physik.unizh.ch}
\affiliation{Physik-Institut der Universit\"{a}t Z\"{u}rich, Winterthurerstrasse 190, CH-8057, Switzerland}

\begin{abstract}

We perform a detailed analysis of the irreversible magnetization
data of Salem-Sugui \textit{et al}. and Bab\'{\i}c \textit{et al}.
of underdoped and optimally doped YBa$_{2}$Cu$_{3}$O$_{7-\delta }$
single crystals. Near the zero field transition temperature we
observe extended consistency with the properties of the 3D-xy
universality class, even though the attained critical regime is
limited by an inhomogeneity induced finite size effect.
Nevertheless, as $T_{c}$ falls from $93.5$ to $41.5$ K \ the
critical amplitude of the in-plane correlation length $\xi _{ab0}$,
the anisotropy $\gamma =\xi _{ab0}/\xi _{c0}$ and the critical
amplitude of the in-plane penetration depth $\lambda _{ab0}$
increase substantially, while the critical amplitude of the $c$-axis
correlation length $\xi _{c0}$ does not change much. As a
consequence, the correlation volume $V_{corr}^{-}$ increases and the
critical amplitude of the specific heat singularity $A^{-}$
decreases dramatically, while the rise of $\lambda _{ab0}$ reflects
the behavior of the zero temperature counterpart. Conversely,
although $\xi _{ab0}$ and $\lambda _{ab0}$ increase with reduced
$T_{c}$, the ratio $\lambda _{ab0}/\xi _{ab0}^{-}$, corresponding to
the Ginzburg-Landau parameter $\kappa _{ab}$, decreases
substantially and YBa$_{2}$Cu$_{3}$O$_{7-\delta }$ crosses over from
an extreme to a weak type-II superconductor.
\end{abstract}

\pacs{74.25.Bt, 74.25.Ha, 74.40.+k}
\maketitle

Since the discovery of cuprate superconductors, fluctuation
contributions to the specific heat,magnetization, magnetic
penetration depths etc. have been measured in a variety of
compounds. In principle, fluctuation studies yield such important
information as the universality class onto which these
superconductors fall and their effective dimensionality.
\cite{book,parks} Given the universality class, various properties,
e.g. the correlation volume above and below $T_{c}$, are no longer
independent but related in terms of universal coefficients. As a
consequence, the isotope and pressure effects on various properties
are no longer independent.\cite{tsiso} In nearly optimally doped
YBa$_{2}$Cu$_{3}$O$_{7-\delta }$, HgBa$_{2}$CuO$_{4-\delta }$ and
La$_{2-x}$Sr$_{x}$CuO$_{4}$ the occurrence of 3D-xy criticality is
reasonably well established.\cite{hub,babic,tsjh2} However, in the
underdoped regime the anisotropy increases with reduced
$T_{c}$.\cite{tsphsycab} Although reduced dimensionality is
accompanied with enhanced fluctuations the 3D-xy critical regime is
expected to shrink. Nevertheless, consistency with 3D-xy scaling was
observed in magnetization measurements of
YBa$_{2}$Cu$_{3}$O$_{7-\delta }$ down to $T_{c}\simeq 61.4$
K.\cite{babic} On the other hand, measurements of the magnetic
penetration depth uncovered charged critical behavior in
YBa$_{2}$Cu$_{3}$O$_{6.95}$ sample with $T_{c}\simeq 61.4$ K.
\cite{tscharge} Note that $7-\delta =6.95$ is close to the hole
concentration $p\simeq 1/8$ where charge fluctuations are
important.\cite{ando}

Recently, magnetization measurements have been performed on
underdoped YBa$_{2}$Cu$_{3}$O$_{7-\delta }$ single crystals with
$T_{c}\simeq 41.5$ K and $T_{c}\simeq 62$ K by Salem-Sugui
\textit{et al}. \cite{salem,salem2} Here we perform a detailed
analysis of these data. In contrast to previous work
\cite{hub,babic} we do not establish the consistency with the 3D-xy
scaling plots only, but estimate, given the critical exponent of the
correlation lengths, $\nu \simeq 2/3$, the critical amplitudes of
the correlation length, the universal ratios, etc., of the
associated fictitious homogeneous system as well. Indeed, the
universality class to which a given experimental system belongs is
not only characterized by its critical exponents but also by various
critical point amplitude ratios and universal coefficients. This is
achieved by invoking the limiting behavior of the universal scaling
function, allowing to explore the growth of the in-plane and
$c$-axis correlation lengths as $T_{c}$ is approached. The observed
limitations of this growth are traced back to a finite size effect,
whereupon the correlation lengths cannot grow beyond the extent of
the homogenous domains. Clearly, such an analysis does not
discriminate between intrinsic or extrinsic inhomogeneities, but it
provides lower bounds for the extent of the homogenous domains seen
by the relevant fluctuations.

The paper is organized as follows: Next we present a short sketch of
the scaling theory and the universal properties appropriate of
anisotropic extreme type-II superconductor exhibiting in the absence
of an applied magnetic field 3D-xy criticality. On this basis we
analyze the irreversible magnetization data of Salem-Sugui
\textit{et al}. \cite{salem,salem2} and Bab\`{\i}c \textit{et al}.
\cite{babic} We observe close to the zero field $T_{c}$ remarkable
consistency with the scaling and critical properties of a finite
system belonging to the 3D-xy universality class. Indeed, the
homogeneity of the samples turns out to be of finite extent,
preventing the correlation lengths to grow beyond the extent of the
homogeneous regions. Accordingly, the magnetization data does not
provide estimates of the scaling and critical properties only, but
uncovers the spatial extent of the homogeneous regions as well. As
$T_{c}$ falls from $93.5$ to $41.5$ K \ we observe that the critical
amplitude of the in-plane correlation length $\xi _{ab0}$, the
anisotropy $\gamma =\xi _{ab0}/\xi _{c0}$ and the critical amplitude
of the in-plane penetration depth $\lambda _{ab0}$ increase
substantially, while the critical amplitude of the $c$-axis
correlation length does not change much. As a consequence, the
correlation volume $V_{corr}^{-}$ increases and the critical
amplitude of the specific heat singularity $A^{-}$ decreases
dramatically, while the rise of $\lambda _{ab0}$ reflects the
behavior of the zero temperature counterpart. \cite{zimmermann}
Conversely, although $\xi _{ab0}$ and $\lambda _{ab0}$increase with
reduced $T_{c}$, the ratio $\lambda _{ab0}/\xi _{ab0}^{-}$,
corresponding to the Ginzburg-Landau parameter $\kappa _{ab}$,
decreases substantially and YBa$_{2}$Cu$_{3}$O$_{7-\delta }$ crosses
over from an extreme to a weak type-II superconductor. The rise of
the anisotropy with reduced $T_{c}$ is consistent with a previous
magnetic torque study\cite{janossy}. For the extent of the
homogenous domains we derive from the finite size effect in $\xi
_{ab}$ and $\xi _{c}$ the lower bounds $L_{ab}\simeq 367$\AA ,
$L_{c}\simeq 53$ \AA\ and $L_{ab}\simeq 254$ \AA\ , $L_{c}\simeq 53$
\AA\ for the samples of Salem-Sugui \textit{et al}.
\cite{salem,salem2} with $T_{c}\simeq 41.5$ and $T_{c}\simeq 62$ K,
respectively.

To derive the scaling form of the magnetization in the fluctuation
dominated regime we note that the scaling of the magnetic field is
in terms of the number of flux quanta per correlation area. Thus,
when the thermal fluctuations of the order parameter dominate the
singular part of the free
energy per unit volume of a homogeneous system scales as \cite%
{book,parks,tsjh2,ffh,tsda,tshkws,tseuro,tsjh}
\begin{equation}
f_{s}=\frac{Q^{\pm }k_{B}T}{\xi _{ab}^{2}\xi _{c}}G\left( z\right)
=\frac{Q^{\pm }k_{B}T\gamma }{\xi _{ab}^{3}}G\left( z\right) ,\text{
}z=\frac{H\xi _{ab}^{2}}{\Phi _{0}}.  \label{eq1}
\end{equation}
$Q^{\pm }$ is a universal constant and $G^{\pm }\left( z\right) $ a
universal scaling function of its argument, with $G^{\pm }\left(
z=0\right) =1$. $\gamma =\xi _{ab}/\xi _{c}$ denotes the anisotropy,
$\xi _{ab}$ the zero-field in-plane correlation length and $H$ the
magnetic field applied along the $c$-axis. Approaching $T_{c}$ the
in-plane correlation length diverges as
\begin{equation}
\xi _{ab}=\xi _{ab0}^{\pm }\left\vert t\right\vert ^{-\nu },\text{
}t=T/T_{c}-1,\text{ }\pm =sgn(t).  \label{eq2}
\end{equation}
Supposing that 3D-xy fluctuations dominate the critical exponents
are given by\cite{peliasetto}
\begin{equation}
\nu \simeq 0.671\simeq 2/3,\text{ }\alpha =2\nu -3\simeq
-0.013,\text{ }\nu \simeq 0.671,  \label{eq3}
\end{equation}
and there are the universal critical amplitude relations\cite
{book,parks,ffh,tsda,tshkws,peliasetto}
\begin{equation}
\frac{\xi _{ab0}^{-}}{\xi _{ab0}^{+}}=\frac{\xi _{c0}^{-}}{\xi
_{c0}^{+}}\simeq 2.21,\text{ }\frac{Q^{-}}{Q^{+}}\simeq 11.5,\text{
}\frac{A^{+}}{A^{-}}=1.07,  \label{eq4}
\end{equation}
and
\begin{equation}
A^{-}\left( \xi _{ab0}^{-}\right) ^{2}\xi
_{c0}^{-}=\frac{A^{-}\left( \xi _{ab0}^{-}\right) ^{3}}{\gamma
}=\left( R^{-}\right) ^{3},R^{-}\simeq 0.815,  \label{eq5}
\end{equation}
where $A^{\pm }$ is the critical amplitude of the specific heat
singularity, defined as $c=\left( A^{\pm }/\alpha \right) \left\vert
t\right\vert ^{-\alpha }+B$. Furthermore, in the 3D-xy universality
class $T_{c}$, $\xi _{c0}^{-}$ and the critical amplitude of the
in-plane penetration depth $\lambda _{ab0}$ are not independent but
related by the universal relation
\cite{book,parks,ffh,tsda,tshkws,peliasetto},
\begin{equation}
k_{B}T_{c}=\frac{\Phi _{0}^{2}}{16\pi ^{3}}\frac{\xi
_{c0}^{-}}{\lambda _{ab0}^{2}}=\frac{\Phi _{0}^{2}}{16\pi
^{3}}\frac{\xi _{ab0}^{-}}{\gamma \lambda _{ab0}^{2}}.  \label{eq6}
\end{equation}

From the singular part of the free energy per unit volume given by
Eq. (\ref{eq1}) we derive for the magnetization per unit volume
$m=M/V=-\partial f_{s}/\partial H$ the scaling form
\begin{equation}
\frac{m}{TH^{1/2}}=-\frac{Q^{\pm }k_{B}\gamma }{\Phi
_{0}^{3/2}}F\left( z\right) ,\text{ }F^{\pm }\left( z\right)
=z^{-1/2}\frac{dG}{dz}, \label{eq7}
\end{equation}
where
\[ z=\frac{\xi _{ab}^{2}}{aL_{H_{c}}^{2}}=x^{-1/2\nu
}=\frac{\left( \xi _{ab0}^{\pm }\right) ^{2}\left\vert t\right\vert
^{-2\nu }H}{\Phi _{0}}.\]
This scaling form is similar to Prange's
\cite{prange} result for Gaussian fluctuations. More generally, the
existence of the magnetization at $T_{c}$, of the penetration depth
below $T_{c}$ and of the magnetic susceptibility above $T_{c}$ imply
the following asymptotic forms of the scaling
function\cite{book,parks,tsjh2,tseuro,tsjh}

\[Q^{\pm }\left. \frac{1}{\sqrt{z}}\frac{dG}{dz}\right\vert _{z\rightarrow
\infty }=Q^{\pm }c_{\infty }^{\pm },\text{ }Q^{-}\left.
\frac{dG}{dz}\right\vert _{z\rightarrow 0}=Q^{-}c_{0}^{-}\left( \ln
z+c_{1}\right) ,
\]

\begin{equation}
Q^{+}\left. \frac{1}{z}\frac{dG}{dz}\right\vert _{z\rightarrow
0}=Q^{+}c_{0}^{+},  \label{eq8}
\end{equation}
with the universal coefficients \cite{book,tsjh2}
\begin{equation}
Q^{-}c_{0}^{-}\simeq -0.7,\text{ }Q^{+}c_{0}^{+}\simeq 0.9,\text{
}q=Q^{\pm }c_{\infty }^{\pm }\simeq 0.5.  \label{eq9}
\end{equation}
The scaling form (\ref{eq7}) with the limits (\ref{eq8}), together
with the critical exponents (Eq. (\ref{eq3})) and the universal
relations (\ref{eq4}) and (\ref{eq6}) are characteristic properties
of the 3D-xy universality class. Accordingly, a homogeneous extreme
type II superconductor falls into this universality class when these
relations are satisfied. When this is the case the doping dependence
of the non-universal critical properties, such as transition
temperature $T_{c}$, critical amplitudes of correlation lengths $\xi
_{ab0,c0}^{\pm }$, anisotropy $\gamma $, \textit{etc}. can be
determined, while the universal relations are independent of the
doping level.

To determine $T_{c}$ we consider the limit $z\rightarrow \infty $
($x\rightarrow 0$). Here the scaling form (\ref{eq7}) reduces with
Eq. (\ref{eq8}) to
\begin{equation}
\frac{m}{H^{1/2}}=-\frac{k_{B}q}{\Phi _{0}^{3/2}}\gamma T,\text{
}q=Q^{\pm }c_{\infty }^{\pm }.  \label{eq10}
\end{equation}
$Q^{+}c_{\infty }^{+}=Q^{-}c_{\infty }^{-}$ follows from the fact
that $m/\sqrt{H_{c}}$ adopts at the zero-field transition
temperature $T_{c}$ a unique value where the curves $m/\sqrt{H}$
\textit{vs}. $T$ taken at different fields $H$ cross and
$m/H^{1/2}\gamma T_{c}$ adopts the universal value
\begin{equation}
\frac{m}{H^{1/2}T_{c}\gamma }=-\frac{k_{B}q}{\Phi _{0}^{3/2}}.
\label{eq11}
\end{equation}
Furthermore, at $T_{c}$ and in the limit $H\rightarrow 0$ Eq.
(\ref{eq10}) also implies
\begin{equation}
\frac{m}{HT_{c}}=-\frac{k_{B}q\gamma }{\Phi
_{0}^{3/2}}\frac{1}{H^{1/2}}, \label{eq12}
\end{equation}
describing the divergence of the diamagnetic susceptibility at
$T_{c}$ when $H\rightarrow 0$. Accordingly, the location of a
crossing point in $m/\sqrt{H} $ \textit{vs}. $T$ provides an
estimate for the 3D transition temperature and the factor of
proportionality in $m/\left( HT_{c}\right) $ \textit{vs}. $H^{1/2}$
probes the anisotropy $\gamma $.

Given then $T_{c}$ and with that the reduced temperature
$t=T/T_{c}-1$, the dominant fluctuations and their properties can
now be explored by invoking the scaling form (\ref{eq7}).
Considering the plot $M/\left( TH^{1/2}\right) $ \textit{vs}. $bt$
for various fixed magnetic fields $H$, according to Eq. (\ref{eq7})
3D-xy fluctuations are verified when $b$ scales as $b\propto
H^{1/2\nu }$ with $\nu \simeq 2/3$. As YBa$_{2}$Cu$_{3}$O$_{7-\delta
}$\ near optimum doping is concerned, the evidence for this 3D-xy
scaling behavior is well established \cite{hub,babic}. Even though
this is an essential step, much more detailed confirmation of 3D-xy
universality and the doping dependence of characteristic critical
properties can be derived by invoking the limiting forms (\ref{eq8})
of the scaling function. Considering the limit $z\rightarrow 0$, Eq.
(\ref{eq7}) reduces below $T_{c}$ to
\begin{equation}
\frac{m}{T}=-\frac{Q^{-}c_{0}^{-}k_{B}}{\Phi _{0}\xi _{c}^{-}}\left(
\ln \left( \frac{H\left( \xi _{ab}\right) ^{2}}{\Phi _{0}}\right)
+c_{1}\right) , \label{eq13}
\end{equation}
and above $T_{c}$ to
\begin{equation}
\frac{m}{TH}=-\frac{Q^{+}c_{0}^{+}k_{B}\left( \xi _{ab}\right)
^{2}}{\Phi _{0}^{2}\xi _{c}}.  \label{eq14}
\end{equation}
Thus, given the magnetization data of a homogenous
system, attaining the limit $z=H\left( \xi _{ab0}^{\pm }\right)
^{2}\left\vert t\right\vert ^{-2\nu }/\Phi _{0}<<1$, the growth of
$\xi _{ab}$ and $\xi _{c} $ is unlimited and estimates for $\xi
_{c0}^{-}$, $\xi _{ab0}^{+}$ and $\left( \xi _{ab0}^{+}\right)
^{2}/\xi _{c0}^{+}$ can be deduced from
\begin{equation} \left\vert
t\right\vert ^{-2/3}\frac{m}{T}=-\frac{Q^{-}c_{0}^{-}k_{B}}{\Phi
_{0}\xi _{c0}^{-}}\left( \ln \left( \frac{H\left( \xi
_{ab0}^{-}\right) ^{2}}{\Phi _{0}}\right) +\ln \left\vert
t\right\vert ^{-4/3}+c_{1}\right) , \label{eq14a}
\end{equation}
and%
\begin{equation}
\left\vert t\right\vert
^{2/3}\frac{m}{TH}=-\frac{Q^{+}c_{0}^{+}k_{B}\left( \xi
_{ab0}^{+}\right) ^{2}}{\Phi _{0}^{2}\xi _{c0}^{+}},  \label{eq14b}
\end{equation}
given the values for the universal constants $Q^{-}c_{0}^{-}$ and
$Q^{+}c_{0}^{+}$ (Eq. (\ref{eq9})). However, for data taken at fixed
magnetic field the reduction of $z=\left( H\left( \xi
_{ab0}^{+}\right) ^{2}/\Phi _{0}\right) \left\vert t\right\vert
^{-4/3}$ unavoidably implies an increasing reduced temperature $t$
and with that a run away from criticality. Thus, for fixed magnetic
field the window where these limiting forms apply is limited to
intermediate values of the reduced temperature. Extrinsic
limitations arise in the presence of inhomogeneities. In this case
the correlation lengths $\xi _{ab,c}$ cannot grow beyond the lengths
$L_{ab,c}$, set by the respective extent of the homogenous domains.
Indeed, there is the Harris criterion \cite{harris}, which states
that short-range correlated and uncorrelated disorder is irrelevant
at the unperturbed critical point, provided that the specific heat
exponent $\alpha $ is negative. Since in the 3D-XY universality
class $\alpha $ is negative (Eq. (\ref{eq3})), a rounded transition
uncovers a finite size effect \cite{cardy,privman}, where the
correlation lengths $\xi _{ab,c}=\xi _{ab0,c0}^{+}\left\vert
t\right\vert ^{-\nu }$ cannot grow beyond $L_{ab,c}$, the respective
extent of the homogenous domains. Hence, as long as $\xi
_{ab,c}<L_{ab,c}$ the critical properties of the fictitious
homogeneous system can be explored with the aid of Eqs.
(\ref{eq14a}) and (\ref{eq14b}). However, closer to $T_{c}$ the
finite size effect sets in as $\xi _{ab,c}$ approaches $L_{ab,c}$.
When $\xi _{c}<L_{c}$ and $\xi _{ab}$ reaches $L_{ab}$ a finite size
effect appears in the plot $\left\vert t\right\vert ^{-2/3}m/T$
\textit{vs}. $\ln \left\vert t\right\vert ^{-4/3}$ around
ln$\left\vert t_{abp}\right\vert ^{-4/3}$ ($\xi _{ab0}^{-}\left\vert
t_{abp}\right\vert ^{-4/3}=L_{ab}$) as the onset of deviations from
the linear behavior. Even closer to $T_{c}$ where both $\xi _{ab}$
and $\xi _{c}$ attain the respective limiting length, $\left(
m/T\right) $ tends according to Eq. (\ref{eq13}) to diverge as
$f_{0}$ $\left\vert t\right\vert ^{-2/3}$ where $f_{0}=-\left(
Q^{-}c_{0}^{-}k_{B}\right) /\left( \Phi _{0}L_{c}\right) \left( \ln
\left( HL_{ab}^{2}/\Phi _{0}\right) +c_{1}\right) $. Accordingly,
sufficiently extended magnetization data do not provide estimates
for the critical properties of the associated fictitious homogeneous
system only, but uncover the extent of the homogenous domains as
well. As a unique size of the homogeneous domains is unlikely, the
smallest extent will set the scale where the growth of the
respective correlation length starts to deviate from the critical
behavior of a homogenous system. Hence, the analysis of
magnetization data allows to probe the homogeneity of the sample as
well, and provides a lower bound for the extent of the homogenous
domains. Furthermore, when the outlined analysis of magnetization
data uncovers 3D-xy universality, it also implies that the pressure
and isotope effects on the critical properties are not independent
but related in terms of universal relations such as
(\ref{eq4})-(\ref{eq6}) and (\ref{eq10}). As an example, the
universal form (\ref{eq10}) implies that the pressure or isotope
exchange induced changes of magnetization, transition temperature
and anisotropy are related by $\Delta m\left( T_{c}\right) /m\left(
T_{c}\right) =\Delta T_{c}/T_{c}+\Delta \gamma \left( T_{c}\right)
/\gamma \left( T_{c}\right) $.\cite{tsiso} Last but not least, it
provides estimates for the pressure and isotope effects on the
critical amplitudes, the correlation lengths, the transition
temperature, the anisotropy and the extent of the homogeneous
domains.

\begin{figure}[htb]
\includegraphics[width=1.0\linewidth]{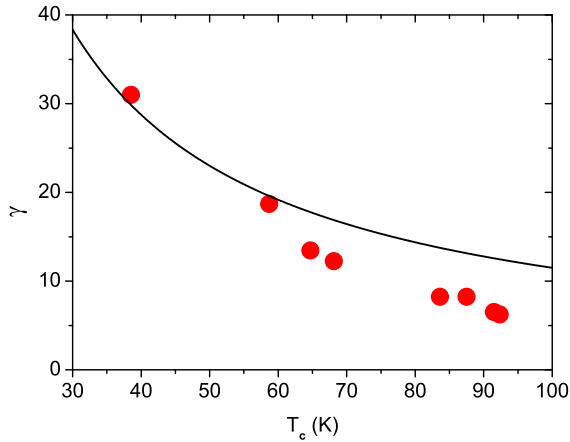}
\vspace{-1.0cm} \caption{$\gamma $ vs. $T_{c}$ for
YBa$_{2}$Cu$_{3}$O$_{7-\delta }$ derived from Janossy \textit{et
al}.\protect\cite{janossy} To estimate $\gamma $ for the $T_{c}$'s
considered here we use $\gamma =1150/T_{c}$ indicated by the solid
line. } \label{fig1}
\end{figure}

We are now prepared to analyze the reversible magnetization data of
Salem-Sugui \textit{et al}. \cite{salem,salem2} for underdoped
YBa$_{2}$Cu$_{3}$O$_{7-\delta }$ single crystals with $T_{c}\simeq
41.5$ K and $T_{c}\simeq 62$ K . From magnetic torque measurements
it is known that in the underdoped regime the chemical substitution
tuned reduction of $T_{c}$ is accompanied by an increase of the
anisotropy. From Fig. \ref{fig1}, showing $\gamma $ vs. $T_{c}$ it
is seen that $\gamma $ increases from $6$ around $T_{c}\simeq 91$ K
to $29$ near $T_{c}\simeq 40$ K.

\begin{figure}[htb]
\includegraphics[width=1.0\linewidth]{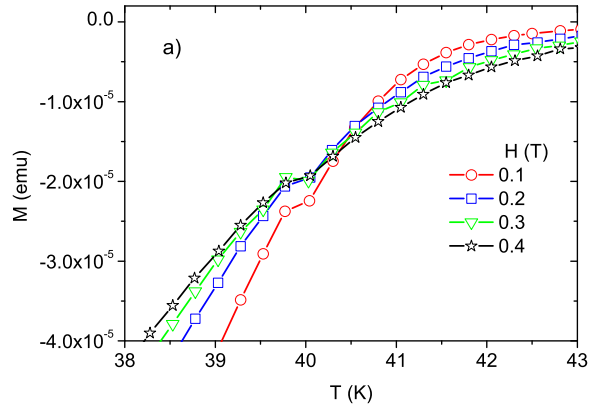}
\includegraphics[width=1.0\linewidth]{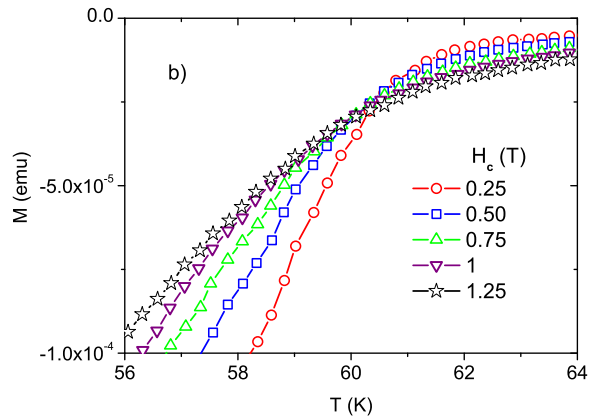}
\vspace{-1.0cm} \caption{Reversible magnetization $M$ \textit{vs}.
$T$ for various fixed magnetic fields $H$ applied along the $c$-axis
of underdoped YBa$_{2}$Cu$_{3} $O$_{7-\delta }$ single crystals with
$T_{c}\simeq 41.5$ K (a) and $T_{c}\simeq 62$ K (b) derived from the
data of Salem-Sugui \textit{et al}. \protect\cite{salem,salem2}
There are nearly field independent crossing points around
$T_{cross}\simeq 40.3$ K and $60.5$ K.} \label{fig2}
\end{figure}

It is well documented that in highly anisotropic cuprate
superconductors the magnetization curves $M(T)$ taken at varying
fields cross below $T_{c}$ around ($T_{cross}$,$M_{cross}$), nearly
independent of the magnitude of the applied field \cite{junod}.
Although this crossing phenomenon is not associated with
criticality, it uncovers that the system behaves below
$T_{cross}$nearly quasi two dimensional \cite{book,junod,huse}. In
Fig. \ref{fig2} \ we depicted the data of Salem-Sugui \textit{et
al}.\cite{salem,salem2} for underdoped YBa$_{2}$Cu$_{3}$O$_{7-\delta
}$ single crystals with $T_{c}\simeq 41.5$ K and $T_{c}\simeq 62$ K
in terms of $M$ \textit{vs}. $T$ for varying $H$ applied along the
$c$-axis. Apparently there are nearly field independent crossing
points around $T_{cross}\simeq 40.3$ K and $60.5$ K. This phenomenon
is however absent in nearly optimally doped and with that
considerably less anisotropic YBa$_{2}$Cu$_{3}$O$_{7-\delta }$
samples. \cite{junod} Since $\gamma $ increases with reduced $T_{c}$
(see Fig. \ref{fig1}), the occurrence of a nearly field independent
crossing point uncovers then a 3D to quasi 2D crossover below
$T_{cross}$.

\begin{figure}[htb]
\includegraphics[width=1.0\linewidth]{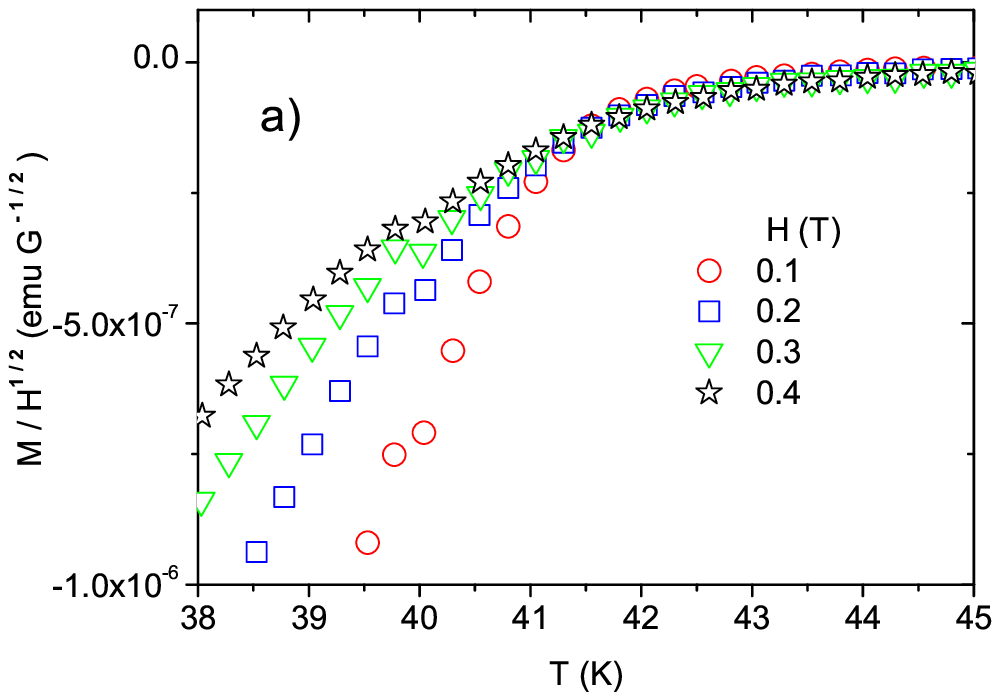}
\includegraphics[width=1.0\linewidth]{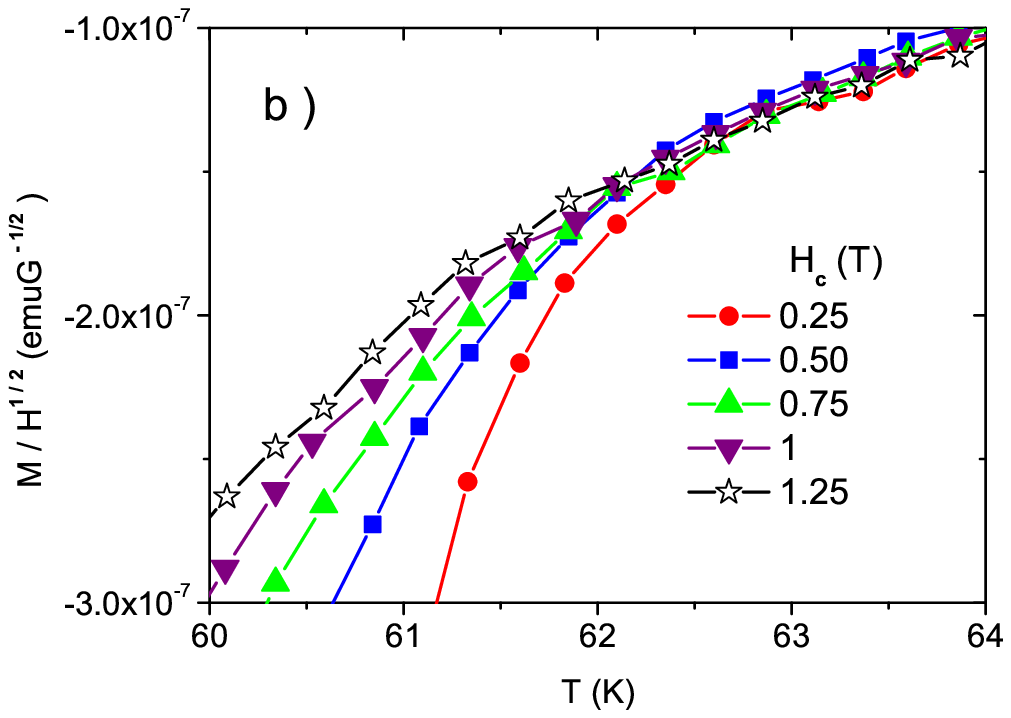}
\vspace{-1.0cm} \caption{$M/H^{1/2}$ \textit{vs}. $T$ at various
fixed magnetic fields $H$ applied along the $c$-axis of underdoped
YBa$_{2}$Cu$_{3}$O$_{7-\delta }$ single crystals with $T_{c}\simeq
41.5$ K (a) and $T_{c}\simeq 62$ K (b) derived from the data of
Salem-Sugui \textit{et al}. \protect\cite{salem,salem2} The crossing
points provide an estimate for the respective transition
temperature, $T_{c}\simeq 41.5$ K (a) and $T_{c}\simeq 62$ K (b)}
\label{fig3}
\end{figure}

Nevertheless, above $T_{cross}$ 3D-xy fluctuations are expected to
set in and to dominate. In this case Eq. (\ref{eq10}) implies the
occurrence of a crossing point in $M/(TH^{1/2})$ \textit{vs}. $T$ at
$T_{c}$. The plots shown in Fig. \ref{fig3} uncover these crossing
points and provide for the respective transition temperatures the
estimates $T_{c}\simeq 41.5$ K and $T_{c}\simeq 62$ K.

\begin{figure}[htb]
\includegraphics[width=1.0\linewidth]{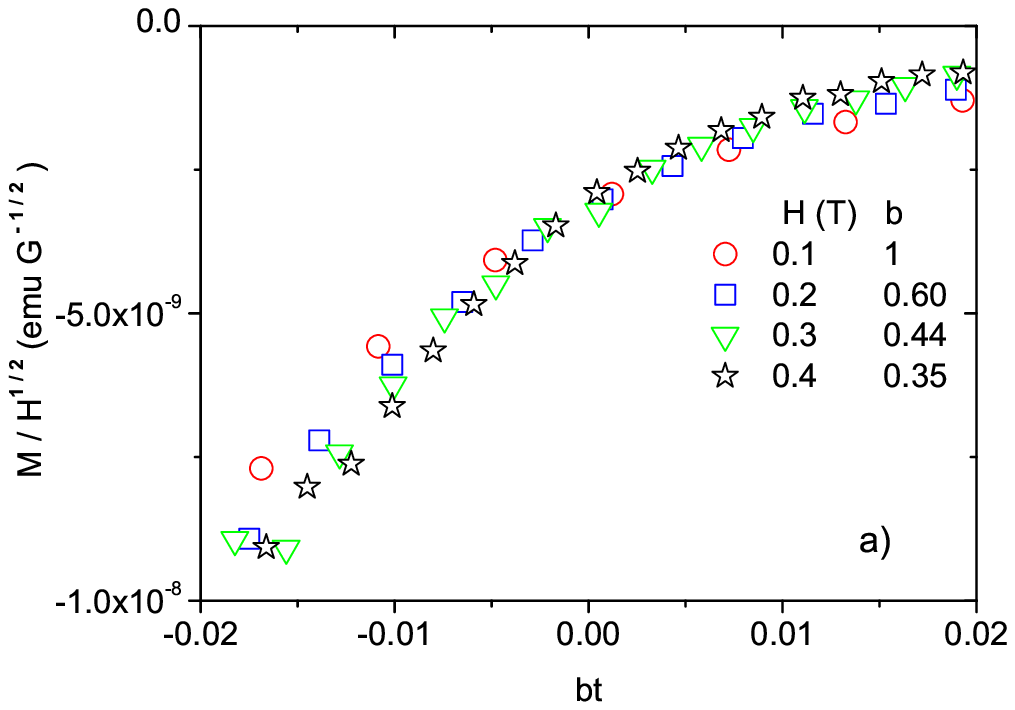}
\includegraphics[width=1.0\linewidth]{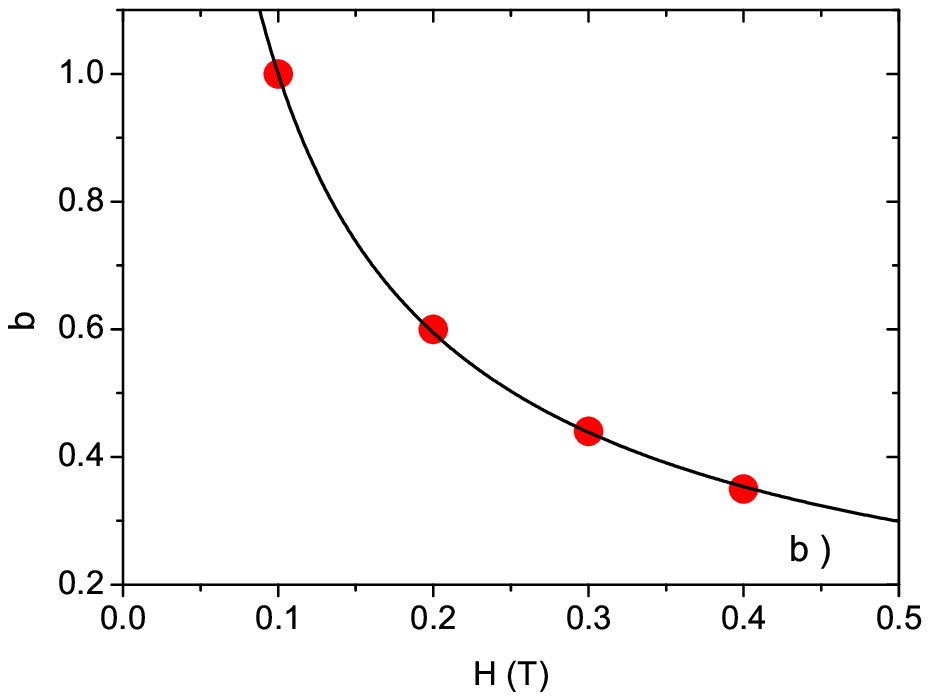}
\vspace{-1.0cm} \caption{a) $M/\left( TH^{1/2}\right) $ \textit{vs}.
$bt$ with $b$ adjusted to achieve a collapse on the $H=0.1$ T curve
for the data shown in Fig. \ref{fig3}a. b) $b$ vs. $H$. The solid
line is $b\propto H^{-3/4}$ characteristic for 3D-xy thermal
fluctuations.} \label{fig4}
\end{figure}

To identify the dominant fluctuations we replotted the data depicted
in Fig. \ref{fig4}a in terms of $M/\left( TH^{1/2}\right) $
\textit{vs}. $bt$ with $b $ adjusted to achieve a collapse on the
$H=0.1$ T curve close to $t=0$. According to Eq. (\ref{eq7}) the
dominance of 3D-xy fluctuations is established when $b$ scales as
$b\propto 1/H^{1/2\nu }\propto 1/H^{3/4}$. In Fig. \ref{fig4}b,
depicting the field dependence of $b$, we observe remarkable
consistency with the characteristic 3D-xy behavior. However, the
quality of the data collapse is seen to deteriorate with increasing
field.

\begin{figure}[htb]
\includegraphics[width=1.0\linewidth]{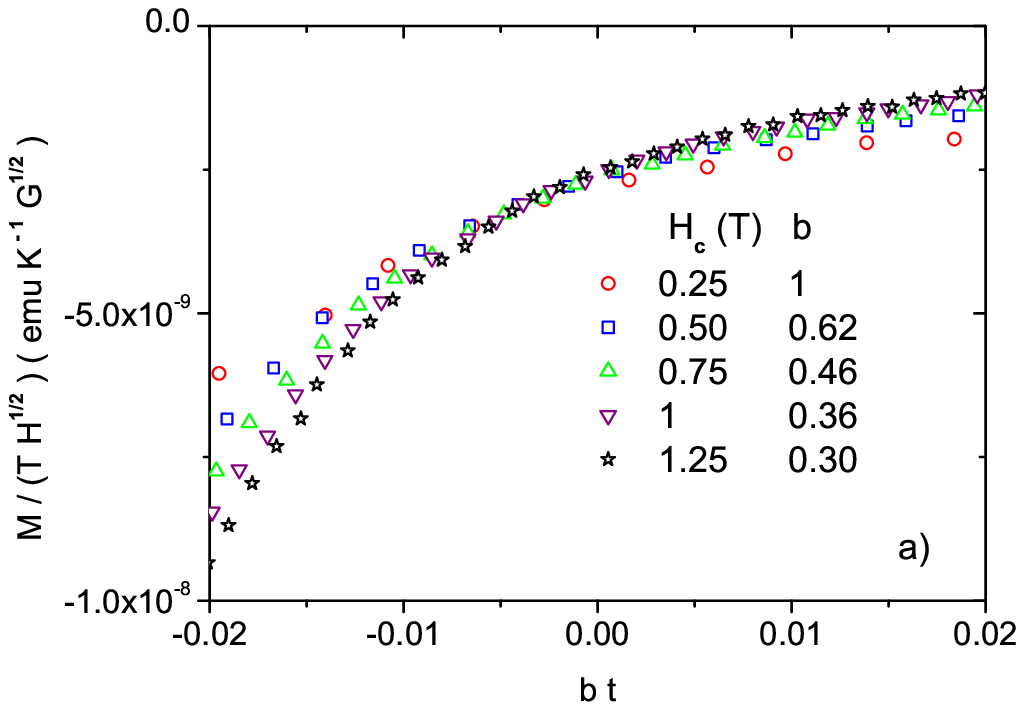}
\includegraphics[width=1.0\linewidth]{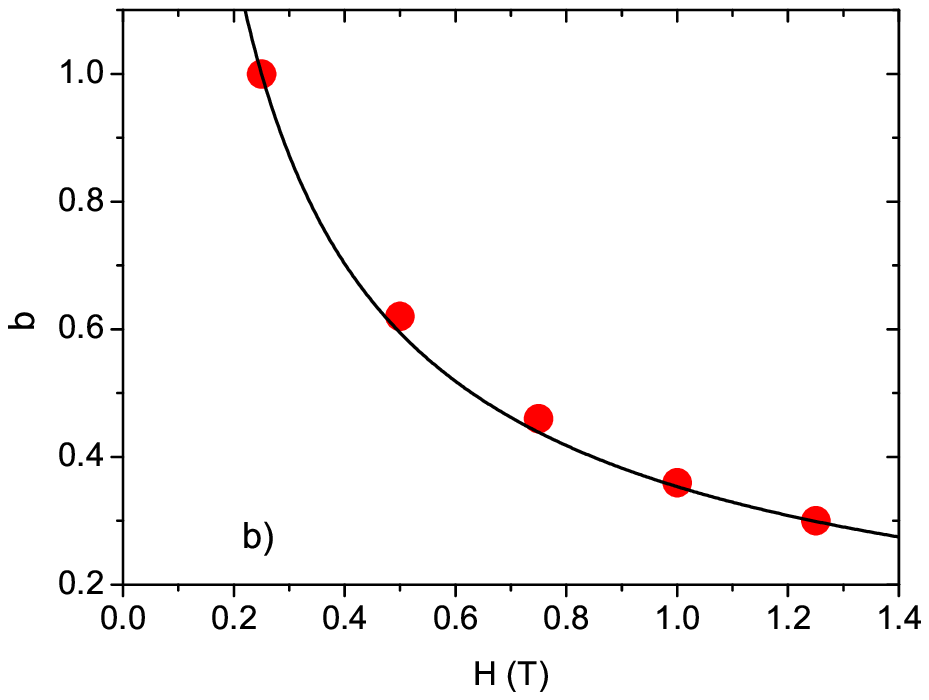}
\vspace{-1.0cm} \caption{a) $M/\left( TH^{1/2}\right) $ \textit{vs}.
$bt$ with $b$ adjusted to achieve a collapse on the $H=0.25$ T curve
for the data shown in Fig. \ref{fig3}b. b) $b$ vs. $H$. The solid
line is $b\propto H^{-3/4}$ characteristic for 3D-xy thermal
fluctuations.} \label{fig5}
\end{figure}

According to Fig. \ref{fig5}, showing the corresponding plots for
the sample with $T_{c}\simeq 62$ K, we observe again consistency
with the characteristic 3D-xy critical behavior, $b\simeq H^{-3/4}$.
Even though the data collapse is seen to deteriorate with increasing
field as well, these plots uncover the dominance of 3D-xy
fluctuations around the estimated transition temperature $T_{c}$.

Having established the consistency with 3D-xy universality in terms
of scaling plots based on the universal scaling form (\ref{eq7})
with $\nu \simeq 2/3$ and the estimates for $T_{c}$ we are no
prepared to perform a more detailed analysis providing additional
checks, allowing to explore the inhomogeneity induced finite size
effects and to estimate the essential critical properties, including
the critical amplitudes of the correlation lengths and the
anisotropy. The basic starting point is either the limiting behavior
below or above $T_{c}$, given by Eqs.(\ref{eq14a}) and
(\ref{eq14b}), respectively. To invoke these limiting forms it is
necessary to convert the magnetization data given in emu to m in emu
cm$^{-3}$ according to Table \ref{Table1}.

\begin{table}
\caption{$T_{c}$, weight of the samples and relationship between $M$
in emu and $m=M\rho /$weight with $\rho \simeq $ $6.3$ g/cm$^{3}$. }
\label{Table1}
\begin{ruledtabular}
\begin{tabular}{lcr}
$T_{c}$ (K) & weight (mg) & $m$ (emu cm$^{-3}$) \\
\hline
41.5 & 1 & 6300 $M$ \\
62 & 1.2 & 5250 $M$ \\
\end{tabular}
\end{ruledtabular}
\end{table}

\begin{figure}[htb]
\includegraphics[width=1.0\linewidth]{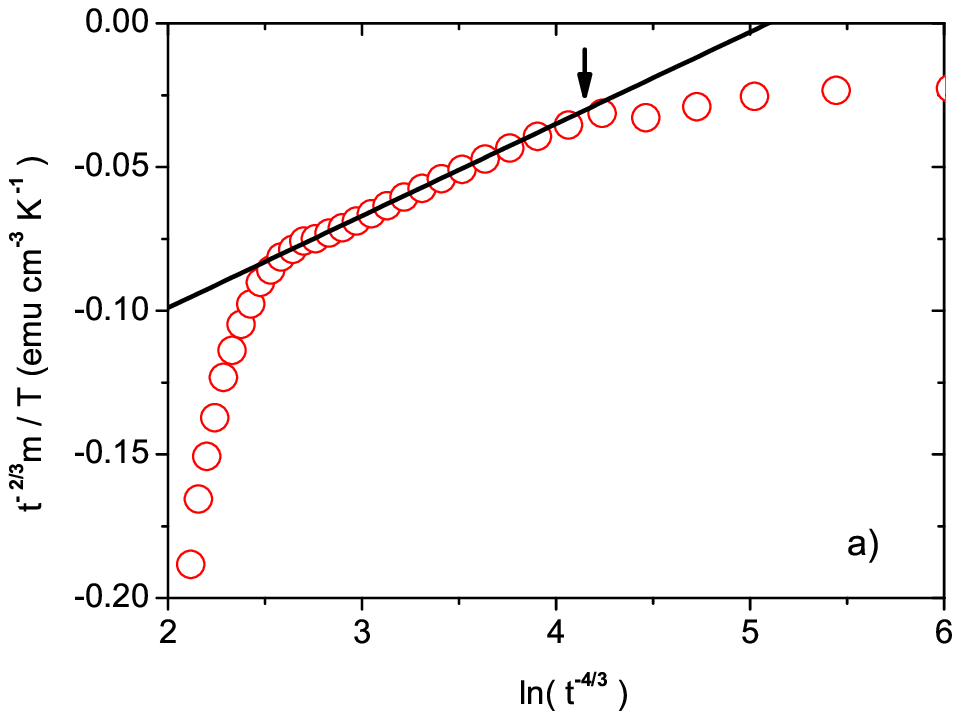}
\includegraphics[width=1.0\linewidth]{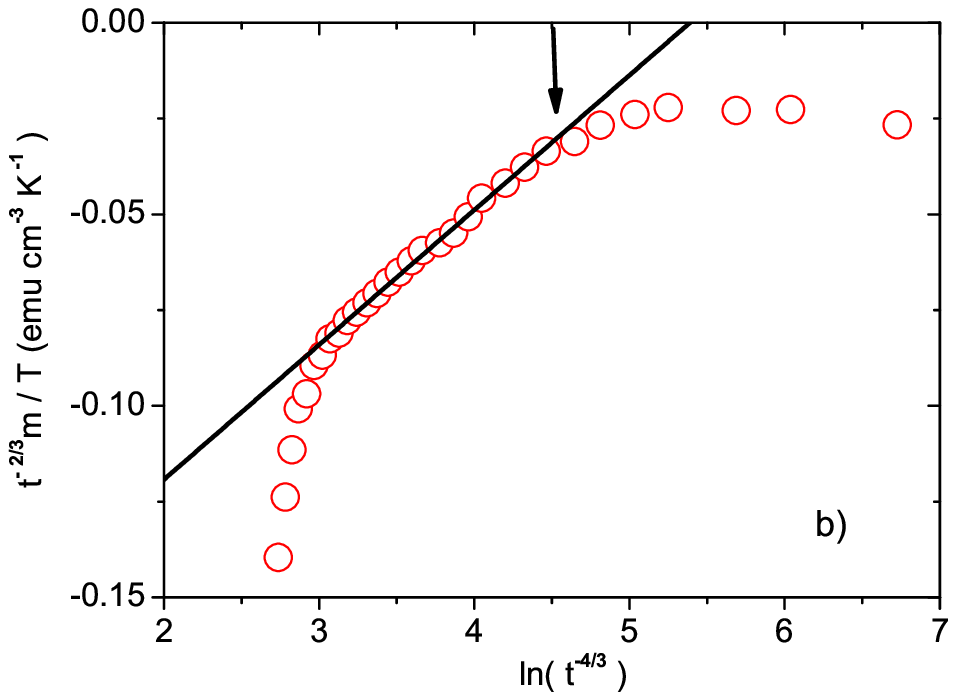}
\vspace{-1.0cm}
\caption{$\left\vert t\right\vert ^{-2/3}m/T$
\textit{vs}. $\ln \left\vert t\right\vert ^{-4/3}$ for $t<0$. a)
Sample with $T_{c}\simeq 41.5$ K at $H=0.1$T; the solid line is
$\left\vert t\right\vert ^{-2/3}m/T=-0.163+0.032\ln \left\vert
t\right\vert ^{-4/3}$. The arrow marks $\ln \left\vert
t_{abp}\right\vert ^{-4/3}\simeq 4.12$ ($t_{abp}\simeq -0.045 $). b)
Sample with $T_{c}\simeq 62$ K at $H=0.25$T; the solid line is
$\left\vert t\right\vert ^{-2/3}m/T=-0.189+0.035\ln \left\vert
t\right\vert ^{-4/3}$. The arrow marks $\ln \left\vert
t_{abp}\right\vert ^{-4/3}\simeq 4.46$ ($t_{abp}\simeq -0.035$).}
\label{fig6}
\end{figure}

Next we invoke the limiting form Eq. (\ref{eq14a}) to estimate the
magnitude of the critical amplitudes below $T_{c}$ and to explore
the homogeneity of the samples. For this purpose we depicted in Fig.
\ref{fig6} $\left\vert t\right\vert ^{-2/3}m/T$ vs. $\ln \left\vert
t\right\vert ^{-4/3}$. In both samples we observe in a limited
interval consistency with the asymptotic behavior, indicated by the
straight lines. Using Eqs. (\ref{eq9}) and (\ref{eq14a}) we derive
from the slope of these lines the estimates
\begin{equation}
\xi _{c0}^{-}\simeq 1.46\text{ \AA\ (}T_{c}\simeq 41.5\text{ K),
}\xi _{c0}^{-}\simeq 1.33\text{ \AA\ (}T_{c}\simeq 62\text{ K),}
\label{eq15}
\end{equation}
for the critical amplitude of the c-axis correlation length.

From the straight lines in Fig. \ref{fig6} it also follows that
\[\ln \frac{10^{3}\left( \xi _{ab0}^{-}\right) ^{2}}{\Phi _{0}}=-5.09-c_{1}\text{ (}T_{c}\simeq 41.5K\text{),}\]
\begin{equation}
\text{ }\ln \frac{2.5\text{ }10^{3}\left( \xi _{ab0}^{-}\right)
^{2}}{\Phi _{0}}=-5.39-c_{1}\text{ (}T_{c}\simeq 62\text{),}
\label{eq16}
\end{equation}
yielding for the critical amplitudes of the in-plane correlation
lengths the ratio
\begin{equation}
\frac{\xi _{ab0}^{-}\left( T_{c}\simeq 41.5\text{ K}\right) }{\xi
_{ab0}^{-}\left( T_{c}\simeq 62\text{ K}\right) }\simeq 1.54.
\label{eq17}
\end{equation}
In addition, this plot reveals that the attainable critical regime,
indicated by the straight lines, ends around the marked values $\ln
\left\vert t_{abp}\right\vert ^{-4/3}\simeq 4.12$ and $\ln
\left\vert t_{abp}\right\vert ^{-4/3}\simeq 4.46$. Here the in-plane
correlation length $\xi _{ab}$ reaches the limiting length $L_{ab}$,
so that $\xi _{ab0}^{-}\left\vert t_{abp}\right\vert
^{-2/3}=L_{ab}$. From $\ln \left\vert t_{abp}\right\vert
^{-4/3}\simeq 4.12$ and $\ln \left\vert t_{abp}\right\vert
^{-4/3}\simeq 4.46$ and Eq. (\ref{eq17}) we obtain for the
ratio between the limiting lengths $L_{ab}$ the estimate%
\begin{equation}
\frac{L_{ab}\left( T_{c}\simeq 41.5\text{ K}\right) }{L_{ab}\left(
T_{c}\simeq 62\text{ K}\right) }\simeq 1.36.  \label{eq20}
\end{equation}
Nevertheless, given the estimate for the universal coefficient
$c_{1}$, we can extract from Eq. (\ref{eq16}) the critical amplitude
of the in-plane correlation length of the respective fictitious
homogeneous system. Using $c_{1}=1.76$, which will be derived later
on, we obtain
\begin{equation}
\xi _{ab0}^{-}\simeq 46.82\text{ \AA : }T_{c}\simeq 41.5\text{ K,
}\xi _{ab0}^{-}\simeq 27.32\text{\AA : }T_{c}\simeq 62\text{ K}
\label{eq21}
\end{equation}
and together with the values for $\xi _{c0}^{-}$ (Eq. (\ref{eq15}))
for the
anisotropy%
\begin{equation}
\gamma \simeq 32.07\text{: }T_{c}\simeq 41.5\text{ K, }\gamma \simeq
20.5\text{: }T_{c}\simeq 62\text{ K,}  \label{eq22}
\end{equation}
in reasonable agreement with the estimates derived from the magnetic
torque measurements, namely $\gamma \simeq 28$ and $19$ (see Fig.
\ref{fig1}).

\begin{figure}[htb]
\includegraphics[width=1.0\linewidth]{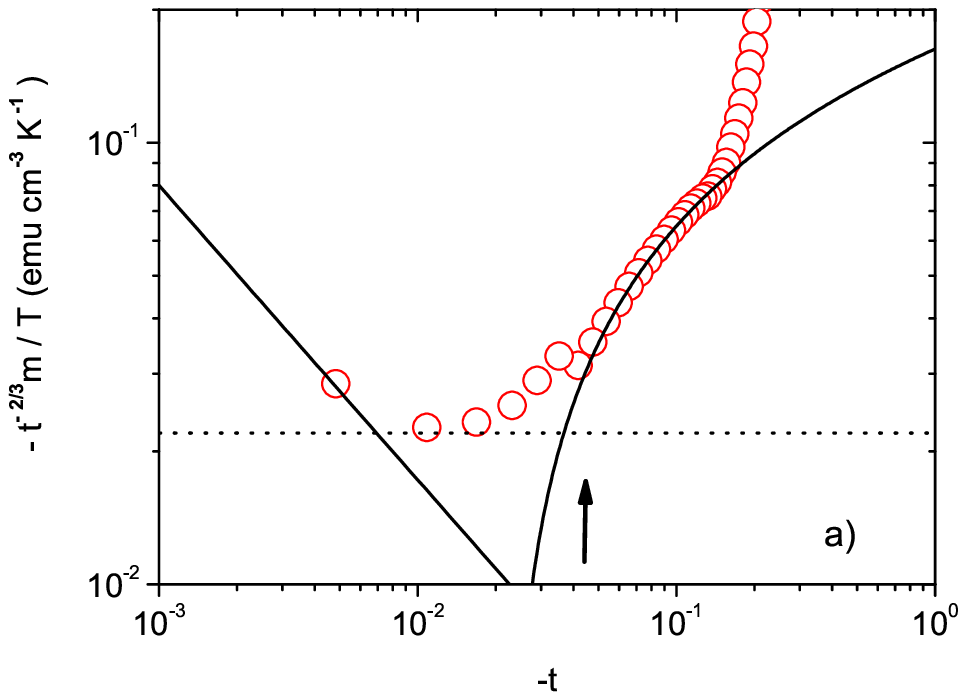}
\includegraphics[width=1.0\linewidth]{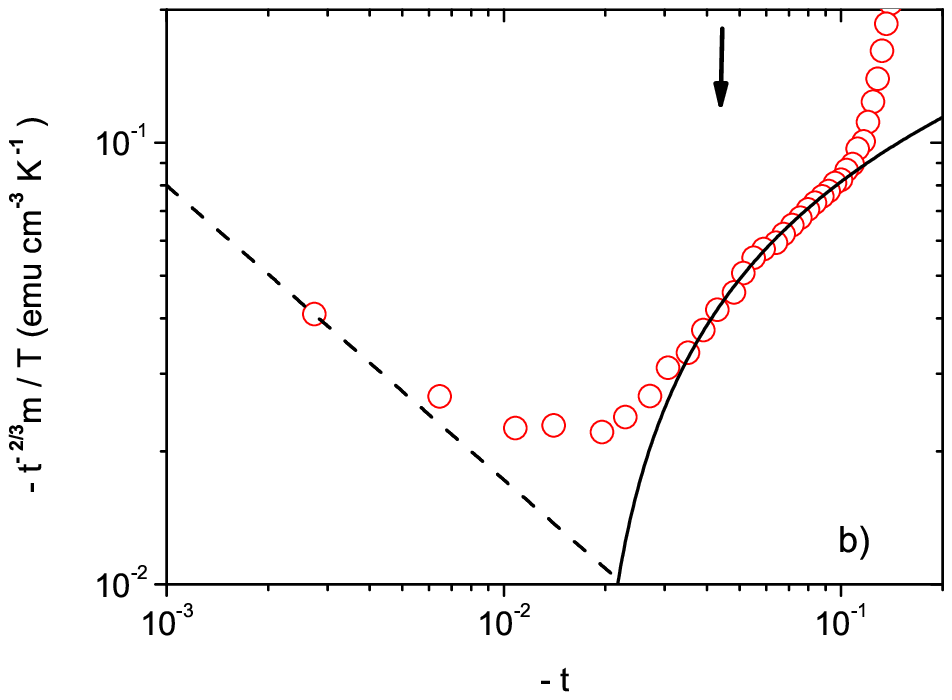}
\vspace{-1.0cm}
\caption{$-\left\vert t\right\vert ^{-2/3}m/T$
\textit{vs}. $-t$ for $t<0$. a) Sample with $T_{c}\simeq 41.5$ K at
$H=0.1$T; the solid curve is $-\left\vert t\right\vert
^{-2/3}m/T=0.15-0.03\ln \left\vert t\right\vert ^{-4/3}$ and the
dashed one $-\left\vert t\right\vert ^{-2/3}m/T=8$
$10^{-4}\left\vert t\right\vert ^{-2/3}$. The arrow marks
$t_{abp}\simeq -0.047$ ($\ln \left\vert t_{abp}\right\vert
^{-4/3}\simeq 4.12)$. b) Sample with $T_{c}\simeq 62$ K at $H=0.25$
T; the solid line is $-\left\vert t\right\vert
^{-2/3}m/T=0.19-0.035\ln \left\vert t\right\vert ^{-4/3}$ and the
dashed one $-\left\vert t\right\vert ^{-2/3}m/T=0.0008\left\vert
t\right\vert ^{-2/3}$. The arrow marks $t_{abp}\simeq -0.042$ ($\ln
\left\vert t_{abp}\right\vert ^{-4/3}\simeq 4.46)$. } \label{fig7}
\end{figure}

To analyze the behavior closer to $T_{c}$ we consider the plot
$-\left\vert t\right\vert ^{-2/3}m/T$ \textit{vs}. $-t$ shown in
Fig. \ref{fig7}. Above $t_{abp}\simeq 0.046$ (Fig. \ref{fig7}a) and
$t_{abp}\simeq 0.034$ (Fig. \ref{fig7}b) there is an interval
exhibiting the $\ln \left\vert t\right\vert ^{-4/3}$ behavior of a
homogeneous system, indicated by the solid curve and consistent with
the $\left\vert t\right\vert ^{-2/3}m/T$ \textit{vs}. $\ln
\left\vert t\right\vert ^{-4/3}$ plots shown in Fig. \ref{fig6}.
Indeed, a minimum occurs below $t_{abp}$ followed by an increase.
Although the data is sparse, it indicates a $\left\vert t\right\vert
^{-2/3}$ divergence, associated with a finite size effect in both,
$\xi _{ab}$ and $\xi _{c}$, whereby these correlation lengths cannot
grow beyond $L_{ab}$ and $L_{c}$, respectively. According to Eq.
(\ref{eq14a}) the amplitude of the divergence is given by
\begin{eqnarray}
\left\vert t\right\vert ^{-2/3}\frac{m}{T}
&=&-\frac{Q^{-}c_{0}^{-}k_{B}}{\Phi _{0}L_{c}}\left( \ln \left(
\frac{HL_{ab}^{2}}{\Phi _{0}}\right)
+c_{1}\right) \left\vert t\right\vert ^{-2/3}  \nonumber \\
&=&f_{0}\left\vert t\right\vert ^{-2/3},  \label{eq23}
\end{eqnarray}
and allows to determine $L_{c}$, given the amplitude $f_{0}$,
$L_{ab}$ and $c_{1}$. From $\ln \left\vert t_{abp}\right\vert
^{-4/3}\simeq 4.12$ (Fig. \ref{fig6}a), $\xi _{ab0}^{-}\simeq 46.82$
\AA\ (Eq. (\ref{eq21})) and $\ln \left\vert t_{abp}\right\vert
^{-4/3}\simeq 4.46$, $\xi _{ab0}^{-}\simeq 27.32$ \AA\ we obtain
with $\xi _{ab}\left( t_{abp}\right) =\xi _{ab}^{-}\left\vert
t_{abp}\right\vert ^{-2/3}=L_{ab}$ for the limiting length in the
$ab$-plane the estimate,
\begin{equation}
L_{ab}\simeq 367\text{ \AA : }T_{c}\simeq 41.5\text{ K,
}L_{ab}\simeq 254\text{ \AA : }T_{c}\simeq 62\text{ K,} \label{eq24}
\end{equation}
in reasonable agreement with the estimate for their ratio given by
Eq. (\ref{eq20}). Together with $Q^{-}c_{0}^{-}\simeq -0.7$ (Eq.
(\ref{eq9})), $c_{1}=1.76$ and the rather crude estimates for
$f_{0}$ (see Fig. (\ref{fig7})) Eq. (\ref{eq23}) yields
\begin{equation}
L_{c}\simeq 57\text{ \AA : }T_{c}\simeq 41.5\text{ K, }L_{c}\simeq
53\text{ \AA : }T_{c}\simeq 62\text{ K,}  \label{eq25}
\end{equation}
in comparison with the previous estimates, $L_{ab}\simeq 392$ \AA\
and $L_{c}\simeq 52$ \AA\ for YBa$_{2}$Cu$_{3}$O$_{6.7}$ with
$T_{c}\simeq 59.7$ K, derived from the inhomogeneity induced finite
size effect in the temperature dependence of the in-plane
penetration depth close to criticality.\cite{tsdan}

\begin{figure}[htb]
\includegraphics[width=1.0\linewidth]{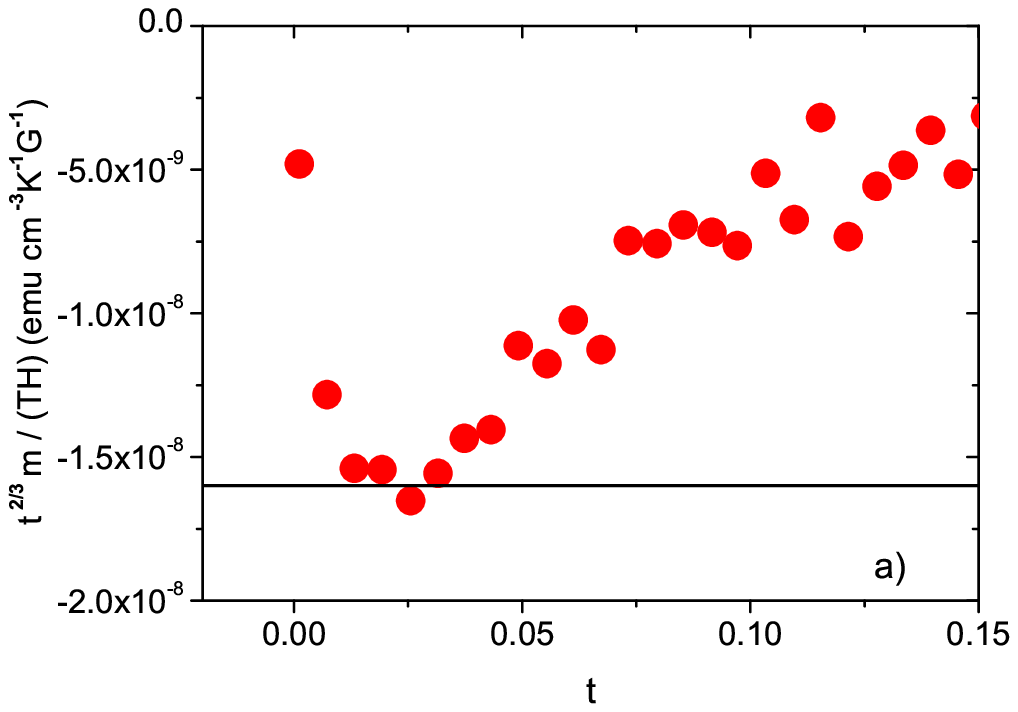}
\includegraphics[width=1.0\linewidth]{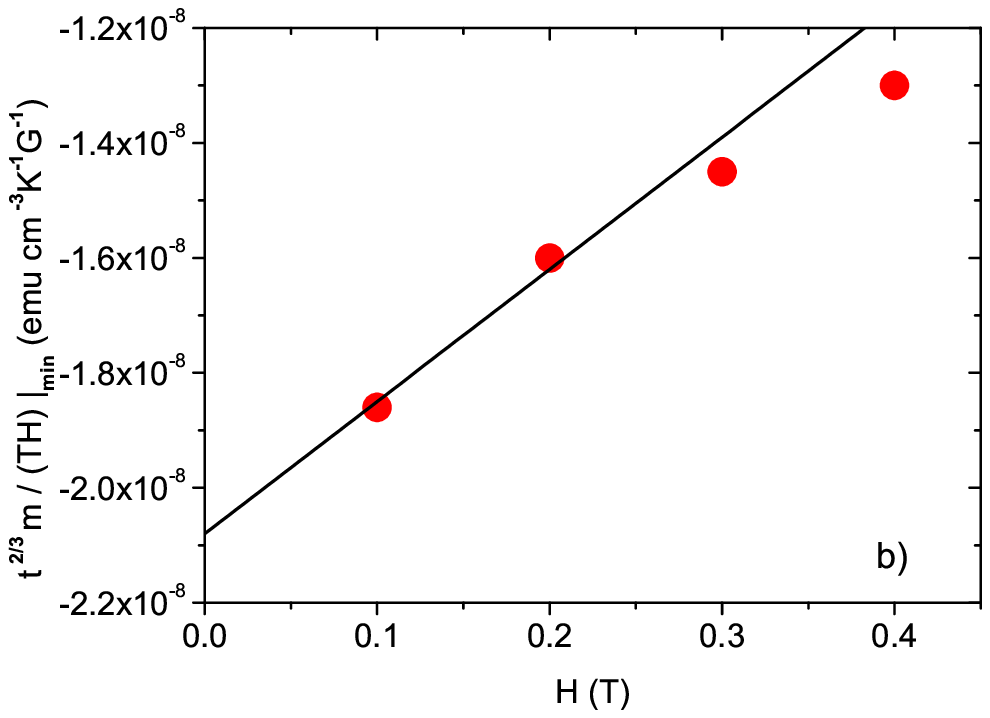}
\vspace{-1.0cm}
\caption{a) $\left\vert t\right\vert ^{2/3}m/\left(
TH\right) $ \textit{vs}. $t $ for the sample with $T_{c}\simeq 41.5$
K and $H=0.2$ T. The horizontal line marks the minimum. b)
$\left\vert t\right\vert ^{2/3}m/\left( TH\right) |_{\min }$
\textit{vs}. $H$ for the sample with $T_{c}\simeq 62$ K and $H=0.25$
T. The line is $\left\vert t\right\vert ^{2/3}m/\left( TH\right)
|_{\min }=-2.08$ $10^{-8}+2.3$ $10^{-8}H$ with $H$ in T.}
\label{fig8}
\end{figure}

Further estimates of the critical amplitudes of the correlation and
limiting lengths can be obtained by invoking Eq. (\ref{eq14}).
According to this we plotted $\left\vert t\right\vert ^{2/3}m/\left(
TH\right) $ \textit{vs}. $t$ for the sample with $T_{c}\simeq 41.5$
K and $H=0.2$ T in Fig. \ref{fig8}a. As $t$ tends to zero the data
approaches a minimum around $t=t_{cp}\simeq 0.026$, indicated by the
horizontal line and marked by the arrow. This minimum provides an
estimate for the critical behavior given by Eq. (\ref{eq14}), while
the upturn below uncovers again an inhomogeneity induced limiting
length along the $c$-axis and in the $ab$-plane. Indeed, when the
growth of $\xi _{ab,c}$ are limited by $L_{ab,c}$, $\left\vert
t\right\vert ^{2/3}m/\left( TH\right) $ tends to zero because the
ratio $\xi _{ab}^{2}/\xi _{c}$ approaches the ratio
$L_{ab}^{2}/L_{c}$. A glance to Fig. \ref{fig8}b reveals that the
minimum exhibits a linear field dependence. Accordingly, the
limiting behavior $dG/dz=c_{0}^{+}z$ (Eq. (\ref{eq8})) is not
attained and the lowest magnetic field dependent correction,
compatible with the linear dependence, $dG/dz=c_{0}^{+}z\left(
1+gz\right) $, must be taken into account. The linear extrapolation
yields
\begin{equation}
\frac{Q^{+}c_{0}^{+}k_{B}\left( \xi _{ab0}^{+}\right) ^{2}}{\Phi
_{0}^{2}\xi _{c0}^{+}}=2.08\text{ 10}^{-8}\text{ (emu
cm}^{-3}\text{K}^{-1}\text{G}^{-1}\text{),}  \label{eq26}
\end{equation}
and with the universal coefficient $Q^{+}c_{0}^{+}\simeq 0.9$
(Eqs.(\ref{eq9}))
\begin{equation}
\left( \xi _{ab0}^{+}\right) ^{2}/\xi _{c0}^{+}=\gamma \xi
_{ab0}^{+}\simeq 718\text{ \AA .}  \label{eq27}
\end{equation}
Invoking then our estimate for the anisotropy, $\gamma \simeq 32.07$
(Eq. (\ref{eq22})) we obtain
\begin{equation}
\xi _{ab0}^{+}\simeq 22.4\text{ \AA , }\xi _{c0}^{+}=\xi
_{ab0}^{+}/\gamma \simeq 0.7\text{ \AA ,}  \label{eq28}
\end{equation}
and with $\xi _{c0}^{-}\simeq 1.46$ \AA\ (Eq. (\ref{eq17})) for the
universal ratio
\begin{equation}
\frac{\xi _{c0}^{-}}{\xi _{c0}^{+}}=\frac{\xi _{ab0}^{-}}{\xi
_{ab0}^{+}}\simeq 2.1,  \label{eq29}
\end{equation}
compared to the theoretical prediction $\xi _{c0}^{-}/\xi
_{c0}^{+}=\xi _{ab0}^{-}/\xi _{ab0}^{+}\simeq 2.21$ (Eq.
(\ref{eq4})). Unfortunately, an equivalent analysis of the data for
the less underdoped sample with $T_{c}\simeq 62$ K is not opportune,
because the measurements do not extend to comparably low fields.

In Table \ref{Table2} we summarized our estimates for the critical
properties of underdoped YBa$_{2}$Cu$_{3}$O$_{7-\delta }$ single
crystals derived from the magnetization data of Salem-Sugui
\textit{et al}. \cite{salem,salem2}. For comparison we included
corresponding values for nearly optimally doped
YBa$_{2}$Cu$_{3}$O$_{7-\delta }$\cite{tseuro}. While the critical
amplitude of the c-axis correlation length $\xi _{c0}^{\pm }$
exhibits a rather weak doping dependence the $ab$-plane counterpart,
$\xi _{ab0}^{\pm }$, and the anisotropy increase drastically with
reduced $T_{c}$. As a consequence, the correlation volume
$V_{corr}^{-}=\left( \xi _{ab0}^{-}\right) ^{2}\xi _{c}^{-}$
increases and the critical amplitude of the specific heat
singularity $A^{-}$ (Eq. (\ref{eq5})) decreases dramatically. This
behavior renders it difficult to extract in the underdoped regime
critical behavior from specific heat data. Because $\xi _{c0}^{-}$
does not change much with reduced $T_{c}$, the critical amplitude of
the in-plane correlation length $\lambda _{ab0}$, resulting from
Eq.~(\ref{eq11}), rises considerably, reflecting the behavior of its
zero temperature counterpart $\lambda _{ab}\left( 0\right) $, where
$\lambda _{ab}\left( 0\right) \simeq 1300$ \AA\ and $\lambda
_{ab}\left( 0\right) \simeq 2250$ \AA\ at $T_{c}=91.3$ K and $60.5$
K, respectively \cite{zimmermann}. Conversely, although both, the
critical amplitudes of the in-plane correlation length $\xi
_{ab0}^{-}$ and penetration depth $\lambda _{ab0}$ increase with
reduced $T_{c}$, the ratio $\lambda _{ab0}/\xi _{ab0}^{-}$,
corresponding to the Ginzburg-Landau parameter $\kappa _{ab}$,
decreases substantially, whereupon YBa$_{2}$Cu$_{3}$O$_{7-\delta }$
crosses over from an extreme to a weak type-II superconductor.
Noting again that the size of the homogeneous domains is not
necessarily unique, the onset of the finite size effect probes their
respective smallest extent. Accordingly our estimates for $L_{ab}$
and $L_{c}$ are lower bounds for the extent of the homogenous
domains.

\begin{table*}[htb]
\caption{Collection of the estimates derived from the magnetization
data of Salem-Sugui \textit{et al}. \cite{salem,salem2} for the
underdoped YBa$_{2}$Cu$_{3}$O$_{7-\delta }$ single crystals. For
comparison we included the estimates for nearly optimally doped
YBa$_{2}$Cu$_{3}$O$_{7-\delta }$.\cite{tseuro} The bracketed values
for $\xi _{c0}^{+}$ and $\xi _{ab0}^{+}$ are obtained with Eq.
(\ref{eq4}). $V_{corr}^{-}=\left( \xi _{ab0}^{-}\right) ^{2}\xi
_{c0}^{-}$ denotes the correlation length volume below $T_{c}$. To
obtain the critical amplitude of the magnetic in-plane penetration
depths $\lambda _{ab0}$ we used the universal relation (\ref{eq6}).}
\label{Table2}
\begin{ruledtabular}
\begin{tabular}{ccccccccccccc}
 $T_{c}$ & $\xi _{c0}^{-}$ & $\xi _{ab0}^{-}$ & $\gamma =\xi
_{ab0}^{-}/\xi _{c0}^{-}$ & $\xi _{c0}^{+}$ & $\left( \xi
_{ab0}^{+}\right) ^{2}/\xi _{c0}^{+}$ & $\xi _{ab0}^{+}$ &
V$_{c}^{-}$ & $L_{ab}$ & $L_{c}$ & 10$^{-3}\lambda _{ab0}$ &
$\lambda _{ab0}/\xi _{ab0}^{-}$ & 10$^{-6}\lambda _{ab0}^{2}$
$T_{c}$ \\
\hline K & \AA  & \AA  &  & \AA  & \AA  & \AA  & \AA
$^{3}$ & \AA  & \AA  & \AA  & & \AA $^{2}$ K \\
\hline
91.7 & 1.3 &10 & 8 & - & - & - & 130 & - & - & 0.941 & 90 & 81.2 \\
62 & 1.33 & 27.32 & 20.5 & (0.60) & - & (12.36) & 993 & 254 & 53 &
1.157
& 42 & 83.0 \\
41.5 & 1.46 & 46.82 &
32.07 & 0.70 & 718 & 22.4 & 3200 & 367 & 57 & 1.482 & 32 & 91.1 \\
\end{tabular}
\end{ruledtabular}
\end{table*}

In this context it is important to recognize that the listed
anisotropies refer to the homogeneous counterparts. However, in the
actual inhomogeneous samples these values only apply in an
intermediate temperature regime where the growth of the correlation
length is not yet limited by the finite size effect. Indeed,
considering the sample with $T_{c}=41.5$ K, below $T_{c}$ $\xi
_{ab}$ levels off around $\left\vert t_{abp}\right\vert \simeq
0.045$, while $\xi _{c}$ saturates around $\left\vert
t_{cp}\right\vert \simeq 0.0041 $. Accordingly, below $\left\vert
t_{abp}\right\vert $ the anisotropy $\gamma =\xi _{ab}/\xi _{c}$
decreases from $\gamma =\xi _{ab0}^{-}/\xi _{c0}^{-}\simeq 32$ to
$\gamma =L_{ab}/L_{c}\simeq 6.4$ at $T_{c}$. This behavior implies
that the 2D-limit in the underdoped regime is hardly accessible
because $\xi _{ab}$ cannot grow beyond $L_{ab}$ and as a result
$\gamma $ does not diverge for fixed $\xi _{c}$.

\begin{figure}[htb]
\includegraphics[width=1.0\linewidth]{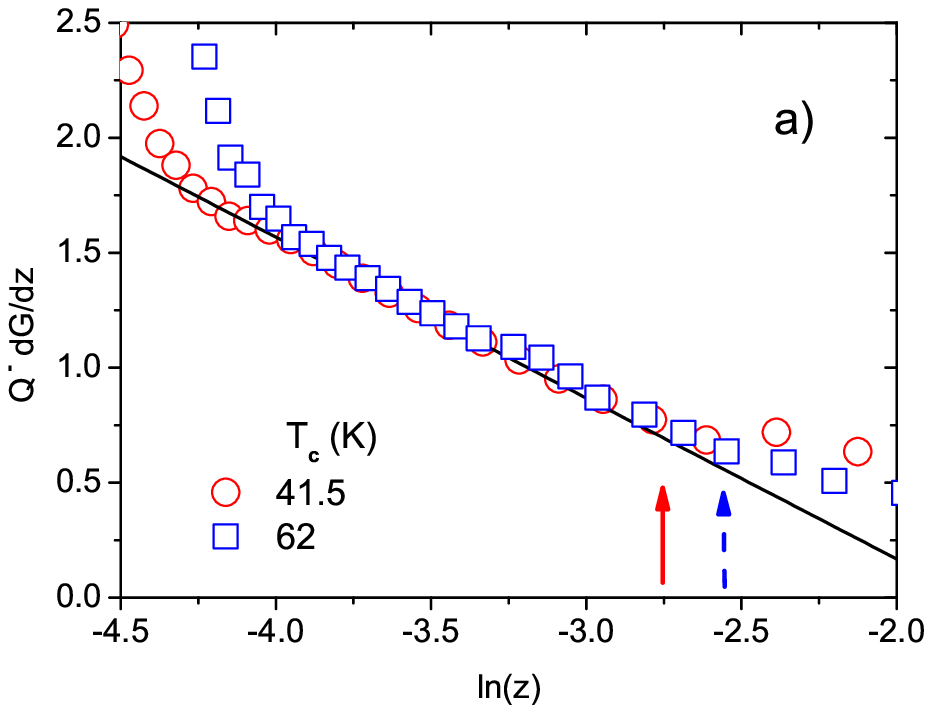}
\includegraphics[width=1.0\linewidth]{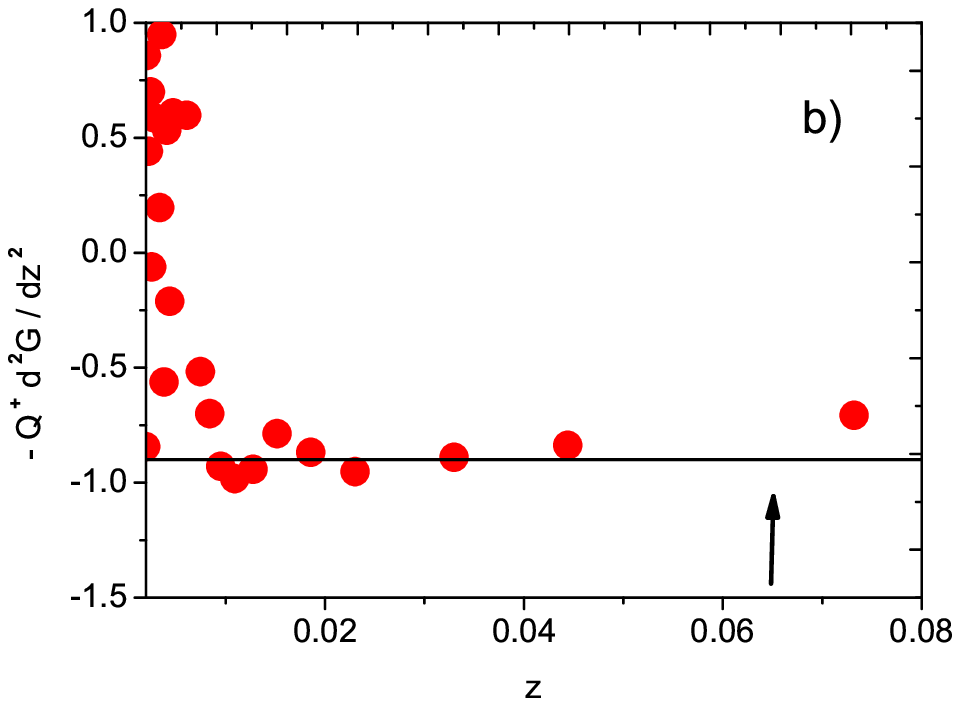}
\vspace{-1.0cm} \caption{ a) $Q^{-}dG/dz=-\left( m\Phi
_{0}/Tk_{B}\right) \xi _{c0}^{-}\left\vert t\right\vert ^{-2/3}$
\textit{vs}. ln$\left( z\right) $ for the sample with $T_{c}=41.5$ K
at $H=0.1$ T, $\xi _{ab0}^{-}=46.82$ \AA\ and $\xi _{c0}^{-}=1.46$
\AA\ ($\bigcirc $) and the sample with $T_{c}=62$ K at $H=0.25$ T,
$\xi _{ab0}^{-}=27.32$ \AA\ and $\xi _{c0}^{-}=1.33$ \AA\ $\left(
\square \right) $; the solid line is $Q^{-}dG/dz$
$=Q^{-}c_{0}^{-}\left( ln\left( z\right) +c_{1}\right) $ with
$z=\left( H\left( \xi _{ab0}^{-}\right) ^{2}/\Phi _{0}\right)
\left\vert t\right\vert ^{-4/3}$, $Q^{-}c_{0}^{-}=-0.7$ (Eq.
(\ref{eq9})), and $c_{1}=1.76$ (Eq. (\ref{eq31})). The arrows mark
the onset of the finite size effect in $\xi _{ab}$ at
ln$(z)=HL_{ab}^{2}/\Phi _{0}$, namely ln$(z)=-2.73$ ($T_{c}=41.5$ K)
and ln$(z)=-2.55$ ($T_{c}=62$ K). b) $Q^{+}d^{2}G/dz^{2}=-d\left(
m\left( \Phi _{0}/Tk_{B}\right) \xi _{c0}^{-}\left\vert t\right\vert
^{-2/3}\right) /dz$ \textit{vs}. $z=\left( H\left( \xi
_{ab0}^{+}\right) ^{2}/\Phi _{0}\right) \left\vert t\right\vert
^{-4/3}$ with $T_{c}=41.5$ K for $\xi _{ab0}^{+}=22.4 $ \AA\ and
$\xi _{c0}^{+}=0.7$ \AA ; the solid line is $Q^{+}d^{2}G/dz^{2}$
$=Q^{+}c_{0}^{+}$ with $Q^{+}c_{0}^{+}\simeq 0.9$ (Eq. (\ref{eq9})).
The arrow marks $z=HL_{ab}^{2}/\Phi _{0}\simeq 0.065$, the onset of
the finite size effect in $\xi _{ab}$.} \label{fig9}
\end{figure}

To check the consistency of our analysis further, we invoke Eq.
(\ref{eq5}) to calculate from the magnetization data the derivative
of the universal scaling function in terms of
\begin{equation}
Q^{\pm }\frac{dG}{dz}=-\frac{m}{T}\frac{\Phi _{0}}{k_{B}}\xi
_{c0}^{\pm }\left\vert t\right\vert ^{-2/3},\text{ }z=\left( H\left(
\xi _{ab0}^{\pm }\right) ^{2}/\Phi _{0}\right) \left\vert
t\right\vert ^{-4/3},  \label{eq30}
\end{equation}
and the respective estimates for the critical amplitudes of the
correlation lengths (Table \ref{Table2}). In Fig. \ref{fig9}a,
showing $-Q^{-}dG/dz$ \textit{vs}. ln$\left( z\right) $ below
$T_{c}$, we observe from ln$\left( z\right) \simeq -2.75$ down to
$-4.25$ consistency with the leading $z\rightarrow 0$ behavior
$Q^{-}dG/dz$ $=Q^{-}c_{0}^{-}\left( ln\left( z\right) +c_{1}\right)
$, where $Q^{-}c_{0}^{-}=-0.7$ (Eq. (\ref{eq9})) and $c_{1}$ was
chosen as
\begin{equation}
c_{1}=1.76.  \label{eq31}
\end{equation}
This estimate fixes the so far unknown universal coefficient
$c_{1}$.The systematic deviations, setting in around ln$\left(
z\right) \simeq -2.75$ ($z\simeq 0.065$), uncover the onset of the
inhomogeneity induced finite size effect, limiting the growth of
$\xi _{ab}$. Indeed, this value corresponds to $z=HL_{ab}^{2}/\Phi
_{0}$ with $H=0.1$T and $L_{ab}=367$ \AA\ (Table \ref{Table2}). On
the other hand, the upturn setting in around ln$(z)\simeq -4.2$
($z\simeq 0.0136$) signals the escape from the scaling regime where
the ln$(z)$ behavior applies. Indeed, in the present case $H$ is
fixed and the reduction of $z=\left( H\left( \xi _{ab0}^{+}\right)
^{2}/\Phi _{0}\right) \left\vert t\right\vert ^{-4/3}$ unavoidably
implies an increasing reduced temperature $t$ and with that a run
away from criticality. Thus, for fixed magnetic field the window
where the universal scaling function can be observed is limited from
below by the run away from criticality and from above by the finite
size effect in $\xi _{ab}$. In Fig. \ref{fig9}b we depicted
$Q^{+}d^{2}G/dz^{2}=-d\left( m\xi _{c0}^{-}\left\vert t\right\vert
^{-2/3}\Phi _{0}/\left( k_{B}T\right) \right) /dz$ \textit{vs}.
$z=\left( H\left( \xi _{ab0}^{+}\right) ^{2}/\Phi _{0}\right)
\left\vert t\right\vert ^{-4/3}$. According to Eq. (\ref{eq9}) it
approaches in a homogenous system in the limit $z\rightarrow 0$ the
universal value $Q^{+}d^{2}G/dz^{2}$ $=Q^{+}c_{0}^{+}\simeq 0.9$.
Even though the available data are rather sparse we observe in the
interval $0.01\lesssim z \lesssim 0.04$ consistency with this
limiting behavior, indicated by the horizontal line. Unfortunately,
the available data do not allow to locate the onset of the finite
size effect in $\xi _{ab}$, seen below $T_{c}$ around ln$\left(
z\right) =\ln \left( HL_{ab}^{2}/\Phi _{0}\right) \simeq \ln \left(
0.065\right) \simeq -2.73$(Fig. \ref{fig9}a). The upturn setting in
around $z=0.01$ signals the runaway from criticality as in
Fig.\ref{fig9}a.

\begin{figure}[htb]
\includegraphics[width=1.0\linewidth]{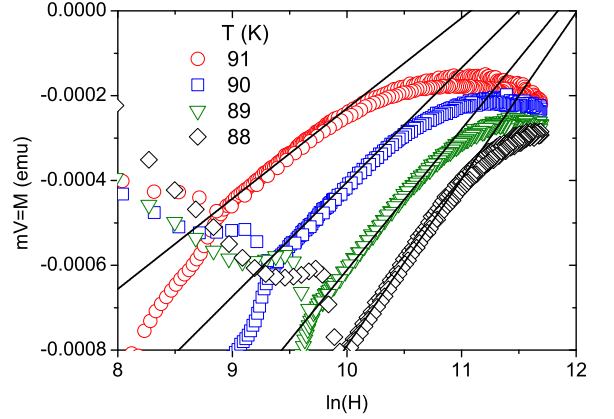}
\vspace{-1.0cm} \caption{$\ M$ \textit{vs}. ln($H$) for various
temperatures derived from the data of Bab\'{\i}c \textit{et al}.
\protect\cite{babic} for a YBa$_{2}$Cu$_{3}$O$_{7-\delta }$ single
crystal with $T_{c}\simeq 93.5$ K and the magnetic field applied
along the $c$-axis. The solid lines are Eq. (\ref{eq32}) with the
parameters listed in Table \ref{Table3}} \label{fig10}
\end{figure}

Next we turn to magnetization data of Bab\'{\i}c \textit{et
al}.\cite{babic}, taken at fixed temperatures below $T_{c}$ as a
function of the magnetic field applied along the $c$-axis. Here we
analyze the data of the YBa$_{2}$Cu$_{3}$O$_{7-\delta }$ single
crystal with $T_{c}\simeq 93.5$ K. In Fig. \ref{fig10} we show $M$
\textit{vs}. ln($H$). In a limited interval we observe linear
behavior so that the scaling form (\ref{eq13}) rewritten in the form
\begin{eqnarray}
M &=&-\frac{VQ^{-}c_{0}^{-}k_{B}T}{\Phi _{0}\xi _{c}}\left( \ln
\left( H\right) +\ln \left( \frac{\xi _{ab}^{2}}{\Phi _{0}}\right)
+c_{1}\right)
\nonumber \\
&=&d+e\ln (H),  \label{eq32}
\end{eqnarray}
applies. The solid lines are this scaling form with the parameters
$d$ and $e $ listed in Table \ref{Table3}.

Given the volume of the sample, the universal amplitudes
$Q^{-}c_{0}^{-}\simeq -0.9$ (Eq. (\ref{eq9})) and $c_{1}$
(Eq.~(\ref{eq31})) the listed correlation lengths are then readily
calculated. Together with $\xi _{ab,c}=\xi _{ab0,c0}^{-}\left\vert
t\right\vert ^{-2/3}$ we obtain for the critical amplitudes the
estimates
\begin{equation}
\xi _{ab0}^{-}\simeq 7\text{ \AA },\text{ }\xi _{c0}^{-}\simeq
1.4\text{ \AA },  \label{eq33}
\end{equation}
in comparison with $\xi _{ab0}^{-}\simeq 10.4$ \AA\ and $\xi
_{c0}^{-}\simeq 1.3$ \AA\ for the sample with $T_{c}=91.7$ K (Table
\ref{Table2}).

\begin{table}[htb]
\caption{Parameters entering Eq. (\ref{eq32}), yielding the straight
lines in Fig. \ref{fig11} and with $Q^{-}c_{0}^{-}=-0.7$ (Eq.
(\ref{eq9})), $c_{1}=1.76$ (Eq. (\ref{eq31})) and $V=8.2$
$10^{-4}cm^{-3}$~ \cite{babic} the estimates for the correlation
lengths $\xi _{ab}$ and $\xi _{c}$.} \label{Table3}
\begin{ruledtabular}
\begin{tabular}{ccccc}
$T$(K) & $d$ (emu) & $e$ (emu) & $\xi _{ab}$(\AA ) & $\xi _{c}$(\AA ) \\
\hline
91 & -2.36 10$^{-3}$ & 2.13 10$^{-4}$ & 74.1 & 16.3 \\
90 & -3.10 10$^{-3}$ & 2.70 10$^{-4}$ & 59.5 & 12.8 \\
89 & -3.93 10$^{-3}$ & 3.32 10$^{-4}$ & 50.7 & 10.3 \\
88 & -4.72 10$^{-3}$ & 3.93 10$^{-4}$ & 46.5 & 8.8 \\
\end{tabular}
\end{ruledtabular}
\end{table}

\begin{figure}[htb]
\includegraphics[width=1.0\linewidth]{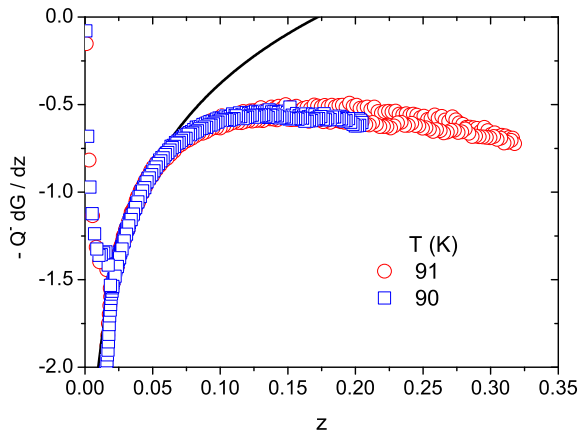}
\vspace{-1.0cm} \caption{$m\Phi _{0}\xi _{c}/\left( k_{B}T\right)
=-Q^{-}dG/dz$ \textit{vs}. $z$ for $T=91$ K and $90$ K. The solid
line is $-Q^{-}dG/dz=-Q^{-}c_{0}^{-}\left( \ln (z+c_{1}\right) $
with $Q^{-}c_{0}^{-}=-0.7$ (Eq. (\ref{eq9})), $c_{1}=1.76$ (Eq.
(\ref{eq31})).} \label{fig11}
\end{figure}

To check the consistency with the scaling form (\ref{eq32}) further,
we calculated $m\Phi _{0}\xi _{c}/\left( k_{B}T\right) =-Q^{-}dG/dz$
\textit{vs}. $z$ for $T=91$ K and $90$ K as shown in Fig.
\ref{fig11} with the parameters listed in Table \ref{Table3}. The
comparison with the leading $z\rightarrow 0$ behavior,
$-Q^{-}dG/dz=-Q^{-}c_{0}^{-}\left( \ln (z+c_{1}\right) $, reveals
that this regime is attained, but limited by irreversibility in the
limit $z\rightarrow 0$ and the crossover to the large $z$-limit ,
$-Q^{-}dG/dz=-qz^{1/2}$ with $q\simeq 0.5$ (Eq. (\ref{eq9})). This
limitation is also apparent in Fig. \ref{fig10} on the low and high
field side. Indeed, because the data do not extend close to $T_{c}$,
the correlation lengths are comparatively small and their growths is
not yet limited by the extent of the homogenous domains.

Although the analyzed magnetization data is afflicted with
uncertainties arising from the subtraction of the normal state
paramagnetism and the Curie term due to paramagnetic impurities or
defects, we observed remarkable consistency with 3D-xy critical
behavior for both, nearly optimally doped and underdoped samples. In
contrast to previous work \cite{hub,babic} we did not establish the
consistency with the 3D-xy scaling plots only, but estimated, given
the critical exponent of the correlation lengths, $\nu \simeq 2/3$,
the critical amplitudes of the correlation length, the universal
ratios, etc., of the associated fictitious homogeneous system as
well. Indeed, the universality class to which a given experimental
system belongs is not only characterized by its critical exponents
but also by various critical point amplitude ratios and universal
coefficients. This has been achieved by invoking the limiting
behavior of the universal scaling function $dG/dz$, allowing via
Eqs. (\ref{eq13})-(\ref{eq14b}) to explore the growth of the
in-plane and $c$-axis correlation lengths as $T_{c}$ is approached
from below or above. We have seen that this growth is limited due to
inhomogeneities and that this limitation appears to be equivalent to
a finite size effect, whereupon the correlation lengths cannot grow
beyond the extent of the homogenous domains. Clearly, such an
analysis does not discriminate between intrinsic or extrinsic
inhomogeneities, but it provides lower bounds for the extent of the
homogenous domains seen by the relevant fluctuations. Even though
since the discovery of superconductivity in the cuprates by Bednorz
and M\"{u}ller\cite{bed} a tremendous amount of work has been
devoted to their characterization, the issue of inhomogeneities and
their characterization is still a controversial issue. There is
neutron spectroscopic evidence for nanoscale cluster formation and
percolative superconductivity in various
cuprates\cite{mesot,furrer}. Nanoscale spatial variations in the
electronic characteristics have been observed in underdoped
Bi$_{2}$Sr$_{2}$CaCu$_{2}$O$_{8+\delta }$ with scanning tunneling
microscopy\cite{liu,chang,cren,lang}. They reveal a spatial
segregation of the electronic structure into $3$-nm-diameter
superconducting domains in an electronically distinct background.
Furthermore, the investigations of Gauzzi \textit{et
al}.\cite{gauzzi} on YBa$_{2}$Cu$_{3}$O$_{6.9}$ films with reduced
long-range structural order clearly reveals that the size of the
homogeneous domains strongly depends on the growth conditions. In
any case we have shown that the analysis of reversible magnetization
data taken near criticality does not uncover the critical properties
of the associated fictitious homogenous and infinite system only,
but provides lower bounds for the extent of the homogeneous domains
as well. Last but not least, having established the consistency with
3D-xy universality there are universal relations such as
(\ref{eq4})-(\ref{eq6}) and (\ref{eq10}). They imply that the effect
of pressure and isotope exchange on the respective properties are
not independent.

\bigskip

The author is grateful to S. Salem-Sugui Jr. and J. R. Cooper for
providing the magnetization data.


\begin{thebibliography}{99}
\bibitem{book} T. Schneider and J. M. Singer, \textit{Phase Transition
Approach To High Temperature Superconductivity}, (Imperial College
Press, London, 2000).

\bibitem{parks} T. Schneider, in: \textit{The Physics of Superconductors},
edited by K. Bennemann and J. B. Ketterson (Springer, Berlin,
(2004), p. 111.

\bibitem{tsiso} T. Schneider, Phys. Rev. B \textbf{67}, 134514 (2003).

\bibitem{hub} M. A. Hubbard,M. B. Salamon, and B. W. Veal, Physica C \textbf{259}, 309 (1996).

\bibitem{babic} D. Babic, J. R. Cooper, J. W. Hodby, and Chen Changkang,
Phys. Rev. B \textbf{60}, 698 (1999).

\bibitem{tsjh2} J. Hofer, T. Schneider, J. M. Singer, M. Willemin, H.
Keller, T. Sasagawa, K. Kishio, K. Conder, and J. Karpinski, Rev. B
\textbf{62}, 631 (2000).

\bibitem{tsphsycab} T. Schneider, Physica B \textbf{326}, 289 (2003).

\bibitem{tscharge} T Schneider, R. Khasanov, K. Conder, E Pomjakushina, R
Bruetsch, and H Keller, J. Phys.: Condens. Matter 16, L 437 (2004).

\bibitem{ando} Y. Ando and K. Segawa, Phys. Rev. Lett. \textbf{88},167005
(2002).

\bibitem{salem} S. Salem-Sugui Jr.,A. D. Alvarenga, K. C. Goretta, V. N.
Vieira, B. Veal, and A. P. Paulikas, Journ. Low Temp. Phys.,
\textbf{141}, 83 (2005).

\bibitem{salem2} S. Salem-Sugui Jr., A. D. Alvarenga, B. Veal, and A. P.
Paulikas, J. Low. Temp. Phys. to be published.

\bibitem{zimmermann} P. Zimmermann \textit{et al}., Phys. Rev. B \textbf{52}, 541 (1995).

\bibitem{janossy} B. Janossy, D. Prost, S. Pekker, and L. Fruchter, Physica
C \textbf{181},51(1991).

\bibitem{ffh} D. S. Fisher, M. P. A. Fisher and D. A. Huse, Phys. Rev. B
\textbf{43}, 130 (1991).

\bibitem{tsda} T. Schneider and D. Ariosa, Z. Phys. B \textbf{89}, 267
(1992).

\bibitem{tshkws} T. Schneider and H. Keller, Int. J. Mod. Phys. B \textbf{8}, 487 (1993).

\bibitem{tseuro} T. Schneider, J. Hofer, M. Willemin, J.M. Singer, and H.
Keller, Eur. Phys. J. B \textbf{3}, 413 (1998).

\bibitem{tsjh} J. Hofer, T. Schneider, J. M. Singer, M. Willemin, H. Keller,
C. Rossel, and J. Karpinski, Pys. Rev. B \textbf{60}, 1332 (1999).

\bibitem{peliasetto} A. Peliasetto and E. Vicari, Physics Reports \textbf{368}, 549 (2002).


\bibitem{prange} R. E. Prange, Phys. Rev. B \textbf{1}, 2349 (1970).

\bibitem{harris} A. B. Harris, J. Phys. C \textbf{7}, 1671 (1974).

\bibitem{cardy} J. L. Cardy ed., \textit{Finite-Size Scaling}, North
Holland, Amsterdam 1988.

\bibitem{privman} V. Privman, \textit{Finite Size Scaling and Numerical
Simulations of Statistical Systems, World Scientific}, NJ, 1990.

\bibitem{junod} A. Junod, J. Y. Genoud, G. Triscone, and T. Schneider,
Physica C \textbf{294}, 115 (1998).

\bibitem{huse} V. Oganesyan, D. A. Huse, and S. L. Sondhi, Phys. Rev. B
\textbf{73}, 094503 (2006).

\bibitem{harshman} D. R. Harshman and A. P. Millis, Phys. Rev. B \textbf{45}, 10684 (1992).

\bibitem{tsdan} T. Schneider and D. Di Castro, Phys. Rev. B \textbf{69},
024502 (2004).

\bibitem{bed} G. Bednorz and K.A. M\"{u}ller, Z. Phys. B: Condens. Matter
\textbf{64}, 189 (1986).

\bibitem{mesot} J. Mesot, P. Allensbach, U. Staub, and A. Furrer, Phys. Rev.
Lett. \textbf{70}, 865 (1993).

\bibitem{furrer} A. Furrer \textit{et al}., Physica C \textbf{235-240}, 261
(1994).

\bibitem{liu} J. Liu, J. Wan, A. Goldman, Y. Chang, and P. Jiang, Phys. Rev.
Lett. \textbf{67}, 2195 (1991).

\bibitem{chang} A. Chang, Z. Rong, Y. Ivanchenko, F. Lu, and E. Wolf, Phys.
Rev. B \textbf{46}, 5692 (1992).

\bibitem{cren} T. Cren, D. Roditchev, W. Sacks, J. Klein, J.-B. Moussy, C.
Deville-Cavellin, and M. Lagu\"{e}s, Phys. Rev. Lett. \textbf{84},
147 (2000).

\bibitem{lang} K.M. Lang, V. Madhavan, J.E. Hoffman, E.W. Hudson, H. Eisaki,
S. Uchida, and J.C. Davis, Nature (London) \textbf{415}, 413 (2002).

\bibitem{gauzzi} A. Gauzzi \textit{et al}., Europhys. Lett. \textbf{51},
667, (2000).

\end{thebibliography}
\end{document}